\documentclass[a4paper,10pt]{revtex4}
\usepackage{mathrsfs}
\usepackage{graphicx}
\usepackage{latexsym}
\usepackage{amsmath}
\usepackage{amssymb}
\usepackage{textcomp}
\usepackage{amsbsy}
\usepackage{graphics}
\usepackage{epstopdf}
\usepackage{color}

\allowdisplaybreaks[4]
\begin{document}

\title{Barrow entropic dark energy: A member of generalized holographic dark energy family}

\author{Shin'ichi~Nojiri$^{1,2}$\,\thanks{nojiri@gravity.phys.nagoya-u.ac.jp},
Sergei~D.~Odintsov$^{3,4}$\,\thanks{odintsov@ieec.uab.es},
Tanmoy~Paul$^{5,6}$\,\thanks{pul.tnmy9@gmail.com}} \affiliation{
$^{1)}$ Department of Physics, Nagoya University,
Nagoya 464-8602, Japan \\
$^{2)}$ Kobayashi-Maskawa Institute for the Origin of Particles
and the Universe, Nagoya University, Nagoya 464-8602, Japan \\
$^{3)}$ ICREA, Passeig Luis Companys, 23, 08010 Barcelona, Spain\\
$^{4)}$ Institute of Space Sciences (IEEC-CSIC) C. Can Magrans
s/n, 08193 Barcelona, Spain\\
$^{5)}$ Department of Physics, Chandernagore College, Hooghly - 712 136.\\
$^{6)}$ International Laboratory for Theoretical Cosmology, TUSUR, 634050 Tomsk, Russia}

\begin{abstract}
The holographic cut-off, in the formalism of generalized holographic dark energy (HDE), is generalized to depend on 
$L_\mathrm{IR} = L_\mathrm{IR} \left( 
L_\mathrm{p}, \dot L_\mathrm{p}, 
\ddot L_\mathrm{p}, \cdots, L_\mathrm{f}, \dot L_\mathrm{f}, \cdots, a\right)$, 
where $L_\mathrm{p}$ and $L_\mathrm{f}$ are the particle horizon and 
future horizon respectively, and $a$ is the scale factor of the universe. Based on such formalism, we showed that the Barrow entropic dark energy (DE) 
model is equivalent to the generalized HDE where the respective holographic cut-off is determined by two ways -- (1) in terms of particle horizon and its 
derivative and (2) in terms of future horizon and its derivative. Interestingly, such cut-off turns out to depend up-to first 
order derivative of $L_\mathrm{p}$ or $L_\mathrm{f}$ respectively. Such equivalence between the Barrow entropic dark energy and the generalized HDE 
is extended to the scenario where the exponent of the Barrow entropy allows to vary with the cosmological expansion of the universe. In both the cases (whether the Barrow exponent 
is a constant or varies with the cosmological evolution), we determine effective equation of state (EoS) parameter from the generalized holographic point 
of view, which, by comparing with the Barrow DE EoS parameter, further ensures the equivalence between the Barrow entropic dark energy 
and the generalized HDE.
\end{abstract}

\maketitle

\section{Introduction}
The holographic principle originates from black hole thermodynamics and string theory and establishes a connection 
of the infrared cutoff of a quantum field theory, which is related to the vacuum energy, with the largest distance of 
this theory \cite{tHooft:1993dmi,Susskind:1994vu,Witten:1998qj,Bousso:2002ju}. 
Such holographic consideration is extensively applied in the field of cosmology, especially 
for the description of late time dark energy era, popularly known as holographic dark energy (HDE) 
\cite{Li:2004rb,Li:2011sd,Wang:2016och,Pavon:2005yx,Nojiri:2005pu,Enqvist:2004xv,Zhang:2005yz,Guberina:2005fb,
Ito:2004qi,Gong:2004cb,BouhmadiLopez:2011xi,Malekjani:2012bw,Khurshudyan:2016gmb,Landim:2015hqa,
Gao:2007ep,Li:2009bn,Feng:2007wn,Lu:2009iv,Huang:2004wt,Mukherjee:2017oom,Nojiri:2021iko,
Bolanos:2021wmn,Cruz:2018lcx,Granda:2018biv,Saridakis:2017rdo,Bouhmadi-Lopez:2017kvc,
Chattopadhyay:2016mnl,Albarran:2015tga,Chakraborty:2020tge,Chakraborty:2020jsq,Chimento:2012zz}. 
Beide the description of dark energy era, the holographic principle also earned a considerable success in the context of early universe both for 
inflation and bounce scenario respectively \cite{Horvat:2011wr,Nojiri:2019kkp,Paul:2019hys,Bargach:2019pst,Elizalde:2019jmh,Oliveros:2019rnq,
Nojiri:2020wmh,Nojiri:2019yzg,Brevik:2019mah,Coriano:2019eif}. 
In the formulation of holographic cosmology, 
the holographic energy density arises due to the holographic principle combined with the dimensional analysis. This makes 
the holographic cosmology (whether it is applied to inflation or to dark energy as well) significantly different compared to the usual inflation or 
dark energy models where some suitable scalar field(s) or higher curvature term(s) present in the Lagrangian play the role of respective agent. 

During the early stage, the size of the universe was small and consequently the holographic energy density is large. 
This, in turn, gets able to trigger a viable inflationary scenario during the early universe, 
which further seems to be consistent with the Planck constraints 
\cite{Horvat:2011wr,Nojiri:2019kkp,Paul:2019hys,Bargach:2019pst,Elizalde:2019jmh,Oliveros:2019rnq,Nojiri:2020wmh}. 
Recently we proposed an unified cosmological scenario of universe from inflation to dark energy (with intermediate 
radiation and matter like era) from holographic point of view \cite{Nojiri:2020wmh}. 
Furthermore, the holographic correspondence of $F(R)$ gravity inflation with/without matter 
fields has been established in \cite{Paul:2019hys}. From a different perspective, the bounce universe has been investigated from holographic side, and as 
a result, it turns out that the presence of holographic energy density helps to violate the null energy condition and leads to a non-singular bounce 
\cite{Nojiri:2019yzg,Brevik:2019mah,Coriano:2019eif}. 
Coming to the dark energy (DE) context, the holographic energy density is inversely proportional to the squared of the infrared cut-off. Such cut-off 
is usually considered to be same as the particle horizon or the future horizon, however, the choice of fundamental viable cut-off is still a debatable 
topic. Consequently, the authors of \cite{Nojiri:2005pu} proposed the most general form of infrared cut-off 
($L_\mathrm{IR}$) and formulated a $generalized$ version of HDE (known as ``generalized'' HDE) where 
they considered the cut-off to be a function of particle horizon, future horizon and their derivatives of various orders, 
in particular, $L_\mathrm{IR} = L_\mathrm{IR} \left(L_\mathrm{p}, \dot L_\mathrm{p}, 
\ddot L_\mathrm{p}, \cdots, L_\mathrm{f}, \dot L_\mathrm{f}, \cdots, a\right)$ with $L_\mathrm{p}$ and $L_\mathrm{f}$ being represented the 
particle horizon and the future horizon respectively. This general form of the cut-off provides a more richer phenomenology to the 
generalized HDE compared to that of the usual HDE. 

On other hand, recently the Barrow entropic dark energy model earned a lot of attention due to a viable description 
of the dark energy era of universe \cite{Saridakis:2020zol,Barrow:2020kug,Adhikary:2021xym,Anagnostopoulos:2020ctz,Srivastava:2020cyk,Bhardwaj:2021chg,
Saridakis:2020lrg,Pradhan:2021cbj,Mamon:2020spa,Sarkar:2021izd}. 
The Barrow entropy was proposed in \cite{Barrow:2020tzx} where it has been showed that quantum 
gravitational effects may introduce fractal features on the black hole structures, which may be encoded within the entropy function given by 
$S \propto A^{1+\Delta}$ ($A$ symbolizes the standard horizon area). 
The exponent $\Delta$ has the range of $0 < \Delta < 1$, and for $\Delta = 0$, it resembles with the 
usual Bekenstein-Hawking entropy function. Depending on the value of $\Delta$, the Barrow entropic dark energy equation of state parameter 
leads to a quintessence region or a phantom region or may experience a phantom crossing during the cosmological evolution. 

Coming to the generalized HDE, besides having a rich phenomenology, it also leads to the following question:
\begin{itemize}
\item Does there exist any suitable form of the $L_\mathrm{IR}$, such that the Barrow entropic dark energy model 
can be regarded as equivalent to the generalized HDE ? If so, then what is the form of the equivalent $L_\mathrm{IR}$ corresponding to the Barrow entropic 
DE model ?
\end{itemize}
We intend to address this question in the present paper. In this regard, our investigation will be carried for two cases: (1) when the 
Barrow exponent is a constant and (2) when the Barrow exponent is relaxed to vary with the cosmological expansion of the universe.

The paper is organized as follows: in Sec.~\ref{Sec0}, we briefly describe the thermodynamics of spacetime and the Barrow entropy function. 
Then in the consecutive two sections, we examine the equivalence between the Barrow entropic dark energy (with constant exponent) and the generalized HDE. The case 
where the exponent varies with the cosmological evolution is discussed in Sec.~\ref{Sec4}. Finally the paper ends with some conclusions and future 
directions.

\section{Thermodynamics of Spacetime and Application to Cosmology}\label{Sec0}
After the thermodynamical properties of the black hole were clarified 
\cite{Bekenstein:1974ax,Hawking:1974sw} and it has been claimed that the entropy of the 
black hole is proportional to the area $A$ of the horizon 
\begin{equation}
\label{Tslls5}
S = \frac{A}{4G}\, ,\quad A = 4\pi r_H^2\, ,
\end{equation}
which is called as the Bekenstein-Hawking entropy,  
$r_H$ is the horizon radius, and we work in units where $\hbar=k_B = c = 1$, 
there have been long and active studies where the connection between the gravity 
and the thermodynamics 
could be clarified \cite{Jacobson:1995ab,Padmanabhan:2003gd}. 
In the studies, we have found that the FRW equations can be also 
regarded as the first law of 
thermodynamics when we consider the Bekenstein-Hawking entropy by using 
the cosmological apparent horizon \cite{Cai:2005ra} 
as a realization of the thermodynamics of space-time \cite{Jacobson:1995ab}.

Recently Barrow was inspired by the Covid-19 virus illustrations and he argued 
that quantum-gravitational effects may introduce intricate, fractal features on the black-hole structure. This 
complex structure leads to finite volume but with infinite (or finite) area, which may be encoded within the black hole entropy function. In particular, 
the Barrow entropy is given by \cite{Barrow:2020tzx},
\begin{equation}
\label{Barrow-entropy}
S_\mathrm{B} = \frac{A_0}{4 G} \left(\frac{A}{A_0} \right)^{1+\Delta}\, .
\end{equation}
In the above expression, $A_0$ is a constant and $\Delta$ 
is the parameter that quantifies the quantum gravitational deformation. Clearly for $\Delta = 0$, the Barrow entropy resembles with the Bekenstein-Hawking 
entropy and $\Delta = 1$ indicates the most fractal black hole structure. 
If the Barrow entropy is applied to cosmology, the Friedmann equations get modified and the modifications can be regarded as a source of dark energy 
density \cite{Saridakis:2020zol,Barrow:2020kug,Adhikary:2021xym,Anagnostopoulos:2020ctz,Srivastava:2020cyk,Bhardwaj:2021chg,
Saridakis:2020lrg,Pradhan:2021cbj,Mamon:2020spa,Sarkar:2021izd}.

\section{Dark Energy corresponding to Barrow entropy} \label{SecI}


In the present context, we consider the Friedmann-Lema\^{i}tre-Robertson-Walker (FLRW) metric with flat spatial section, in particular, 
\begin{equation}
ds^2=-dt^2+a^2(t)\sum_{i=1,2,3} \left(dx^i\right)^2 \, ,
\label{metric}
\end{equation}
where $a(t)$ is known as the scale factor of the universe. The Hubble rate is defined by $H=\frac{\dot a}{a}$, and consequently the radius of the 
cosmological horizon (symbolized by $r_h$) is given by, 
\begin{equation}
\label{apphor}
r_H=\frac{1}{H}\, .
\end{equation}
As a result, the entropy inside the cosmological horizon is given by the Bekenstein-Hawking relation. Furthermore 
the energy flux ($E$) or equivalently the amount of increase of heat ($Q$) in the aforementioned horizon (having radius $r_h$) comes as,
\begin{equation}
\label{Barrow2}
dQ = - dE = -\frac{4\pi}{3} r_H^3 \dot\rho dt = -\frac{4\pi}{3H^3} \dot\rho dt 
= \frac{4\pi}{H^2} \left( \rho + p \right) dt \, ,
\end{equation}
where, in the last equality, we use the conservation law corresponding to the matter energy density 
$\rho$, in particular, $\dot \rho + 3 H \left( \rho + p \right) = 0$. By incorporating the first law of thermodynamics, i.e. $TdS = dQ$, into 
Eq.~(\ref{Barrow2}), one gets,
\begin{align}
 T\frac{dS}{dt} = \frac{4\pi}{H^2} \left( \rho + p \right)\, .\label{Barrow-new1}
\end{align}
The above expression along with the Hawking temperature defined by \cite{Cai:2005ra},
\begin{equation}
\label{Barrow6}
T = \frac{1}{2\pi r_H} = \frac{H}{2\pi}\, ,
\end{equation}
lead to the second FLRW equation as,
\begin{align}
\dot H = - 4\pi G \left( \rho + p \right)\, ,
\label{Barrow-new2}
\end{align}
and by integrating both sides of Eq.~(\ref{Barrow-new2}), one obtains the first FLRW equation, 
\begin{equation}
\label{Barrow8}
H^2 = \frac{8\pi G}{3} \rho + \frac{\Lambda}{3} \, ,
\end{equation}
where the integration constant $\Lambda$ acts as a cosmological constant.

Having set the stage, we now apply the above formalism for the case of the Barrow entropy, instead of the Bekenstein-Hawking entropy. 
By incorporating the first law of thermodynamics to the Barrow entropy, the FLRW equations get modified compared to the 
Eqs.~(\ref{Barrow-new2}) and (\ref{Barrow8}) respectively. Using Eq.~(\ref{Barrow-entropy}), we determine,
\begin{align}
\frac{dS}{dt} = \frac{dS}{dA}\frac{dA}{dt} = -\frac{8\pi}{H^3}\left(\frac{1+\Delta}{4G}\right)\left(\frac{H_1^2}{H^2}\right)^{\delta}\dot{H}\, ,
\label{Barrow-new3}
\end{align}
where the constant $H_1$ is defined as, $A_0 = 4\pi/H_1^2$. Consequently, the second FLRW equation corresponding to the Barrow entropy comes as,
\begin{align}
\left(1+\Delta\right)\left(\frac{H_1^2}{H^2}\right)^{\Delta}\dot{H} = - 4\pi G \left( \rho + p \right)\, ,
\label{Barrow-FLRW1}
\end{align}
on integrating which, one gets,
\begin{equation}
\label{Barrow-FLRW2}
\frac{1 + \Delta}{1 - \Delta} H_1^2\left( \frac{H^2}{H_1^2} \right)^{1 - \Delta}
= \frac{8\pi G}{3} \rho + \frac{\Lambda}{3} \, ,
\end{equation}
where the integration constant is symbolized by $\Lambda$. The above two FLRW equations can be equivalently written as,
\begin{align}
\dot{H}=&-4\pi G\left[\left(\rho + p\right) + \left(\rho_\mathrm{B} + p_\mathrm{B}\right)\right]~, \nonumber\\
H^2=&\frac{8\pi G}{3} \left( \rho + \rho_\mathrm{B} \right) 
+ \frac{\Lambda}{3} \, ,
\label{Barrow-FLRW-diff-form}
\end{align}
where $\rho_\mathrm{B}$ and $p_\mathrm{B}$ have the following expressions, 
\begin{align}
\rho_\mathrm{B} = \frac{3}{8\pi G} 
\left( H^2 - \frac{1 + \Delta}{1 - \Delta} H_1^2\left( \frac{H^2}{H_1^2} \right)^{1 - \Delta} 
\right)\, ,\label{rhoB}
\end{align}
and
\begin{align}
p_\mathrm{B} = \frac{\dot{H}}{4\pi G}\left\{\left(1 + \Delta\right)\left(\frac{H_1^2}{H^2}\right)^{\Delta} - 1\right\} 
- \frac{3}{8\pi G} 
\left( H^2 - \frac{1 + \Delta}{1 - \Delta} H_1^2\left( \frac{H^2}{H_1^2} \right)^{1 - \Delta} 
\right)~,
\label{pB}
\end{align}
respectively. Eq.~(\ref{Barrow-FLRW-diff-form}) clearly indicates that $\rho_\mathrm{B}$ and $p_\mathrm{B}$ can be identified 
as effective energy density and pressure correspond to the Barrow entropy. 
It may be observed from Eq.~(\ref{rhoB}) that $\rho_\mathrm{B}$ contains the quadratic as well as lesser than quadratic power of the Hubble parameter (due to 
$0 < \Delta < 1$). With the above expressions of $\rho_\mathrm{B}$ and $p_\mathrm{B}$, we can define 
the equation of state (EoS) parameter for the Barrow entropy as,
\begin{align}
\omega_\mathrm{B} = \frac{p_\mathrm{B}}{\rho_\mathrm{B}} 
= -1 + 2\left(\frac{\dot{H}}{3H^2}\right)\left\{\frac{\left(1+\Delta\right)\left(\frac{H_1^2}{H^2}\right)^{\Delta} - 1}
{1 - \frac{1 + \Delta}{1 - \Delta} \left( \frac{H_1^2}{H^2} \right)^{\Delta}}\right\}\, .
\label{eosB}
\end{align}
It may be checked that the above expression of $\omega_\mathrm{B}$ obeys the conservation equation for the Barrow entropic energy density, i.e.
\begin{align}
\dot{\rho}_\mathrm{B} + 3H\rho_\mathrm{B}\left(1 + \omega_\mathrm{B}\right) = 0\, .
\label{conservation-B}
\end{align}
Here it deserves mentioning that the author of \cite{Saridakis:2020zol} showed that the Barrow entropic energy density can lead to a viable 
dark energy epoch of our present universe, particularly the Barrow dark energy model exhibits more interesting and richer phenomenology comparing to the 
standard scenario. Moreover, Eq.~(\ref{eosB}) argues that the exponent $\Delta$ significantly affects the 
EoS parameter, and depending on the value of $\Delta$, the Barrow dark energy can lie in a quintessence regime, 
in a phantom regime, or may experience a phantom-divide crossing during the cosmological evolution \cite{Saridakis:2020zol}.

\section{Generalized Holographic Dark Energy}\label{SecII}

The holographic energy density, in accordance to the holographic principle, gets proportional to the inverse squared infrared cutoff
$L_\mathrm{IR}$, in particular,
\begin{equation}
\label{basic}
\rho_\mathrm{hol}=\frac{3c^2}{\kappa^2 L^2_\mathrm{IR}}\, ,
\end{equation}
where $c$ is a free parameter and $\kappa^2 = 8\pi G$ with $G$ being the Newton's gravitational constant. Usually, 
the holographic cutoff $L_\mathrm{IR}$ is considered to be equivalent with the particle horizon $L_\mathrm{p}$ or the future event horizon
$L_\mathrm{f}$, as given by,
\begin{equation}
\label{H3}
L_\mathrm{p}\equiv a \int_0^t\frac{dt}{a}\ ,\quad
L_\mathrm{f}\equiv a \int_t^\infty \frac{dt}{a}\, ,
\end{equation}
respectively. Differentiation of both sides of the above expressions (with respect to $t$) immediately lead to the Hubble parameter in terms of 
$L_\mathrm{p}$ and its first derivative or in terms of $L_\mathrm{f}$ and its first derivative, i.e.,
\begin{equation}
\label{HLL}
H \left( L_\mathrm{p} , \dot{L}_\mathrm{p} \right) =
\frac{\dot{L}_\mathrm{p}}{L_\mathrm{p}} - \frac{1}{L_\mathrm{p}}\, , 
\quad 
H(L_\mathrm{f} , \dot{L}_\mathrm{f}) = \frac{\dot{L}_\mathrm{f}}{L_\mathrm{f}} + \frac{1}{L_\mathrm{f}}
\, .
\end{equation}
The holographic cut-off is usually considered to be same as the particle horizon or the future horizon. However a more general form 
of the holographic cut-off was proposed in \cite{Nojiri:2005pu} where the cut-off ($L_\mathrm{IR}$) seems 
to depend on the particle horizon as well as on the future event horizon and their various derivatives, in particular,
\begin{equation}
\label{GeneralLIR}
L_\mathrm{IR} = L_\mathrm{IR} \left( L_\mathrm{p}, \dot L_\mathrm{p}, 
\ddot L_\mathrm{p}, \cdots, L_\mathrm{f}, \dot L_\mathrm{f}, \cdots, a\right)\, .
\end{equation}
The other dependency of $L_\mathrm{IR}$, particularly on the Hubble parameter, Ricci scalar and their derivatives, are 
implicitly encapsulated within either $L_p$ and their derivatives or $L_f$ and their derivatives via Eq.~(\ref{HLL}). 
Such generalized version of HDE (known as ``generalized'' HDE), 
introduced by Nojiri and Odintsov in \cite{Nojiri:2005pu}, leads to interesting phenomenology both from inflation and dark energy perspective 
\cite{Nojiri:2020wmh}. Due to the above expression, a general covariant gravity model like,
\begin{equation}
\label{GeneralAc}
S = \int d^4 \sqrt{-g} F \left( R,R_{\mu\nu} R^{\mu\nu},
R_{\mu\nu\rho\sigma}R^{\mu\nu\rho\sigma}, \Box R, \Box^{-1} R,
\nabla_\mu R \nabla^\mu R, \cdots \right) \, ,
\end{equation}
can be regarded to be equivalent to generalized holographic model for a suitable form of cut-off. 
We will use the above expressions frequently in the rest of the paper. 
Based on Eq.~(\ref{GeneralLIR}), our main goal is to show that Barrow holographic dark energy model is actually equivalent to the generalized holographic 
dark energy where the respective holographic cut-off becomes function of either the particle horizon and its derivatives or the future horizon and 
its derivatives.

By comparing Eqs.~(\ref{rhoB}) and (\ref{basic}), we may argue that the Barrow entropic dark energy is a candidate of 
the generalized holographic dark energy family, where the corresponding holographic 
cutoff (symbolized by $L_\mathrm{B}$) gets the following expression, 
\begin{align}
\label{rhoB2}
\frac{3c^2}{\kappa^2 L^2_\mathrm{B}} 
= \frac{3}{8\pi G} 
\left( \left( \frac{\dot{L}_\mathrm{p}}{L_\mathrm{p}} - \frac{1}{L_\mathrm{p}} \right)^2 
- \frac{1 + \Delta}{1 - \Delta} H_1^2\left( \frac{\left( \frac{\dot{L}_\mathrm{p}}{L_\mathrm{p}} - \frac{1}{L_\mathrm{p}} \right)^2}
{H_1^2} \right)^{1 - \Delta} \right) \, , 
\end{align}
in terms of $L_\mathrm{p}$ and its derivatives, similarly $L_\mathrm{B}$, in terms of the future horizon and its 
derivatives, is given by,
\begin{align}
\label{rhoBfuture horizon}
\frac{3c^2}{\kappa^2 L^2_\mathrm{B}} = \frac{3}{8\pi G} 
\left( \left( \frac{\dot{L}_\mathrm{f}}{L_\mathrm{f}} + \frac{1}{L_\mathrm{f}} \right)^2 
- \frac{1 + \Delta}{1 - \Delta} H_1^2\left( \frac{\left( \frac{\dot{L}_\mathrm{f}}{L_\mathrm{f}} + \frac{1}{L_\mathrm{f}} \right)^2}
{H_1^2} \right)^{1 - \Delta} \right) \, .
\end{align}
To get the above two expressions, we use Eq.~(\ref{HLL}). Beside the first Friedmann equation, we also 
need to establish the equivalence between the equation of state (EoS) parameters of the Barrow dark energy and generalized holographic 
dark energy models for our present purpose. Thereby we now determine the effective EoS parameter corresponding to the 
cut-off $L_\mathrm{B}$ or equivalently to the holographic energy density $\rho_\mathrm{hol}^{(B)} = \frac{3c^2}{\kappa^2L_\mathrm{B}^2}$. 
By using the conservation relation of $\rho_\mathrm{hol}^{(B)}$, we get the desired EoS parameter (symbolized by $W_\mathrm{hol}^\mathrm{(B)}$) as follows,
\begin{align}
 W_\mathrm{hol}^\mathrm{(B)} = -1 + \left(\frac{2}{3HL_\mathrm{B}}\right)\frac{dL_\mathrm{B}}{dt}~,
 \label{eos-holB}
\end{align}
where $L_\mathrm{B}$ is obtained in Eq.~(\ref{rhoB2}) (or Eq.~(\ref{rhoBfuture horizon})) and the superscript `B' 
in the above expression denotes the EoS parameter corresponds to the holographic cut-off $L_\mathrm{B}$. Differentiating both sides of 
Eq.~(\ref{rhoB2}) or Eq.~(\ref{rhoBfuture horizon}), we obtain $W_\mathrm{hol}^\mathrm{(B)}$ in terms of $L_\mathrm{p}$ or $L_\mathrm{f}$ as,
\begin{align}
W_\mathrm{hol}^\mathrm{(B)} = -1 + \frac{2\left(\ddot{L}_\mathrm{p}/L_\mathrm{p} - \left(\dot{L}_\mathrm{p}/L_\mathrm{p}\right)^2 
+ \dot{L}_\mathrm{p}/L_\mathrm{p}^2\right)}{3\left(\frac{\dot{L}_\mathrm{p}}{L_\mathrm{p}} - \frac{1}{L_\mathrm{p}}\right)^{2}}
\left\{\frac{\left(1+\Delta\right)\left[H_1^2\left(\frac{\dot{L}_\mathrm{p}}{L_\mathrm{p}} - \frac{1}{L_\mathrm{p}}\right)^{-2}\right]^{\Delta} - 1}
{1 - \frac{1 + \Delta}{1 - \Delta} \left[H_1^2\left(\frac{\dot{L}_\mathrm{p}}{L_\mathrm{p}} - \frac{1}{L_\mathrm{p}}\right)^{-2}\right]^{\Delta}}\right\}\, ,
\label{eos-hol-ph}
\end{align}
or,
\begin{align}
W_\mathrm{hol}^\mathrm{(B)} = -1 + \frac{2\left(\ddot{L}_\mathrm{f}/L_\mathrm{f} - \left(\dot{L}_\mathrm{f}/L_\mathrm{f}\right)^2 
 - \dot{L}_\mathrm{f}/L_\mathrm{f}^2\right)}{3\left(\frac{\dot{L}_\mathrm{f}}{L_\mathrm{f}} + \frac{1}{L_\mathrm{f}}\right)^{2}}
\left\{\frac{\left(1+\Delta\right)\left[H_1^2\left(\frac{\dot{L}_\mathrm{f}}{L_\mathrm{f}} + \frac{1}{L_\mathrm{f}}\right)^{-2}\right]^{\Delta} - 1}
{1 - \frac{1 + \Delta}{1 - \Delta} \left[H_1^2\left(\frac{\dot{L}_\mathrm{f}}{L_\mathrm{f}}+- \frac{1}{L_\mathrm{f}}\right)^{-2}\right]^{\Delta}}\right\}\, ,
\label{eos-hol-fh}
\end{align}
respectively. Therefore the Friedmann equations from such holographic point of view are given by,
\begin{align}
H^2=&\frac{8\pi G}{3}\rho + \left( \left( \frac{\dot{L}_\mathrm{p}}{L_\mathrm{p}} - \frac{1}{L_\mathrm{p}} \right)^2 
 - \frac{1 + \Delta}{1 - \Delta} H_1^2\left( \frac{\left( \frac{\dot{L}_\mathrm{p}}{L_\mathrm{p}} - \frac{1}{L_\mathrm{p}} \right)^2}
{H_1^2} \right)^{1 - \Delta} \right)\, ,\nonumber\\
0=&\frac{d}{dt}\left(\rho_\mathrm{hol}^{(B)}\right) + \frac{2\rho_\mathrm{hol}^{(B}\left(\ddot{L}_\mathrm{p}/L_\mathrm{p} - \left(\dot{L}_\mathrm{p}/L_\mathrm{p}\right)^2 
+ \dot{L}_\mathrm{p}/L_\mathrm{p}^2\right)}{\left(\frac{\dot{L}_\mathrm{p}}{L_\mathrm{p}} - \frac{1}{L_\mathrm{p}}\right)}
\left\{\frac{\left(1+\Delta\right)\left[H_1^2\left(\frac{\dot{L}_\mathrm{p}}{L_\mathrm{p}} - \frac{1}{L_\mathrm{p}}\right)^{-2}\right]^{\Delta} - 1}
{1 - \frac{1 + \Delta}{1 - \Delta} \left[H_1^2\left(\frac{\dot{L}_\mathrm{p}}{L_\mathrm{p}} - \frac{1}{L_\mathrm{p}}\right)^{-2}\right]^{\Delta}}\right\}\, ,
\label{FRW-hol-ph}
\end{align}
or,
\begin{align}
H^2=&\frac{8\pi G}{3}\rho + \left( \left( \frac{\dot{L}_\mathrm{f}}{L_\mathrm{f}} + \frac{1}{L_\mathrm{f}} \right)^2 
 - \frac{1 + \Delta}{1 - \Delta} H_1^2\left( \frac{\left( \frac{\dot{L}_\mathrm{f}}{L_\mathrm{f}} + \frac{1}{L_\mathrm{f}} \right)^2}
{H_1^2} \right)^{1 - \Delta} \right)\, ,\nonumber\\
0=&\frac{d}{dt}\left(\rho_\mathrm{hol}^{(B}\right) + \frac{2\rho_\mathrm{hol}^{(B)}\left(\ddot{L}_\mathrm{f}/L_\mathrm{f} - \left(\dot{L}_\mathrm{f}/L_\mathrm{f}\right)^2 
 - \dot{L}_\mathrm{f}/L_\mathrm{f}^2\right)}{\left(\frac{\dot{L}_\mathrm{f}}{L_\mathrm{f}} + \frac{1}{L_\mathrm{f}}\right)}
\left\{\frac{\left(1+\Delta\right)\left[H_1^2\left(\frac{\dot{L}_\mathrm{f}}{L_\mathrm{f}} + \frac{1}{L_\mathrm{f}}\right)^{-2}\right]^{\Delta} - 1}
{1 - \frac{1 + \Delta}{1 - \Delta} \left[H_1^2\left(\frac{\dot{L}_\mathrm{f}}{L_\mathrm{f}}+- \frac{1}{L_\mathrm{f}}\right)^{-2}\right]^{\Delta}}\right\}\, ,
\label{FRW-hol-fh}
\end{align}
where the second equation(s) represent the conservation equation for the holographic energy density $\rho_\mathrm{hol}^{(B)} \propto \frac{1}{L_\mathrm{B}^2}$. With the 
above Friedmann equations, the conservation relation for normal matter holds. Owing to such expressions, the density parameters 
for holographic energy density turns out to be,
\begin{align}
\Omega_\mathrm{hol}^{(B)} = \frac{8\pi G}{3H^2}\rho_\mathrm{hol}^{(B)}~~.
\label{density parameter}
\end{align}
With the help of Eq.~(\ref{HLL}), we may compare Eq.~(\ref{eos-hol-ph}) (or Eq.~(\ref{eos-hol-fh})) with Eq.~(\ref{eosB}) to get,
\begin{align}
W_\mathrm{hol}^\mathrm{(B)} \equiv \omega_\mathrm{B}\, ,
\end{align}
that, in turn, establishes the equivalence between the EoS parameters of the Barrow dark energy model and the holographic dark energy model 
coming from the cut-off 
$L_\mathrm{B}$. Furthermore as shown in \cite{Saridakis:2020zol}, depending on the values of the exponent $\Delta$, the Barrow dark energy EoS parameter 
may lead to a quintessence or a phantom kind of dark energy model. 
Such equivalence, along with the fact that the Barrow entropic energy density provides a viable dark energy model, lead to the argument 
that the holographic energy density having the cut-off $L_\mathrm{B}$ is also able to produce a viable dark energy epoch at our current universe. 

Thereby as a whole, the Barrow entropic dark energy model may be regarded as a candidate of the generalized holographic dark energy family, with the 
corresponding holographic cut-off is given by Eq.~(\ref{rhoB2}) (or by Eq.~(\ref{rhoBfuture horizon})).\\

Before moving to the next section, we intend to discuss some characteristics of the present HDE model, in particular, we will investigate whether the HDE 
model under consideration allows turning point in the evolution of Hubble parameter \cite{Colgain:2021beg}. 
The Friedmann equations in presence of holographic dark energy density are shown in Eq.(\ref{FRW-hol-ph}) (or Eq.(\ref{FRW-hol-fh})) where 
$\rho$ represents the energy density for normal matter fields. Considering $\rho = \rho_\mathrm{m} + \rho_\mathrm{r}$, with $\rho_\mathrm{m}$ 
consists of baryons and cold dark matter and $\rho_\mathrm{r}$ is the radiation energy density, Eq.(\ref{FRW-hol-ph}) (or similarly Eq.(\ref{FRW-hol-fh})) 
becomes,
\begin{eqnarray}
 H^2 = \frac{\kappa^2}{3}\left(\rho_\mathrm{m} + \rho_\mathrm{r} + \rho_\mathrm{hol}^{(B)}\right)~~,
 \label{new1}
\end{eqnarray}
where recall, $\rho_\mathrm{hol}^{(B)} \propto \frac{1}{L_\mathrm{B}^2}$ is the Barrow entropic HDE density. 
The conservation equation for different components are given by,
\begin{eqnarray}
 \frac{d\rho_\mathrm{m}}{dt} + 3H\rho_\mathrm{m} = 0~~~~~~~~~~~&,&~~~~~~~~~~~\frac{d\rho_\mathrm{r}}{dt} + 4H\rho_\mathrm{r} = 0~,\nonumber\\
 \frac{d\rho_\mathrm{hol}^{(B)}}{dt} +&3H&\left(\rho_\mathrm{hol}^{(B)} + p_\mathrm{hol}^{(B)}\right) = 0~~,
 \label{new2}
\end{eqnarray}
with $p_\mathrm{hol}^{(B)}$ represents the pressure corresponding to the HDE. Moreover the spatial component of Friedmann equation is,
\begin{eqnarray}
 2\dot{H} + 3H^2 + \kappa^2\left(p_\mathrm{r} + p_\mathrm{hol}^{(B)}\right) = 0~~,
 \label{new3}
\end{eqnarray}
where we use $p_\mathrm{m} = 0$. By combining the Friedmann equations along with the relation $p_\mathrm{r} = \rho_\mathrm{r}/3$, we get 
the pressure for HDE as,
\begin{eqnarray}
 p_\mathrm{hol}^{(B)} = \left(-1 - \frac{2\dot{H}}{3H^2}\right)\rho_\mathrm{T} - \rho_\mathrm{r}/3~~,
 \label{new4}
\end{eqnarray}
with $\rho_\mathrm{T} = \rho_\mathrm{m} + \rho_\mathrm{r} + \rho_\mathrm{hol}^{(B)}$ symbolizes the total energy density. Here, for convenience, we 
define the fractional energy densities of various components like $\Omega_\mathrm{i} = \kappa^2\rho_\mathrm{i}/\left(3H^2\right)$ (the suffix 
'i' runs for different components), and by definition, $\sum \Omega_\mathrm{i} = 1$. Accordingly, the following relation holds true
\begin{eqnarray}
 \frac{d\rho_\mathrm{hol}^{(B)}}{dt} = \rho_\mathrm{T}\left[\frac{d\Omega_\mathrm{hol}^{(B)}}{dt} + \frac{2\dot{H}}{H}\Omega_\mathrm{hol}^{(B)}\right]~~.
 \label{new5}
\end{eqnarray}
Plugging the above expressions of $\dot{\rho}_\mathrm{hol}^{(B)}$ and $p_\mathrm{hol}^{(B)}$ into Eq.(\ref{new2}) yields,
\begin{eqnarray}
 \frac{d\Omega_\mathrm{hol}^{(B)}}{dt} + \frac{2\dot{H}}{H}\left(\Omega_\mathrm{hol}^{(B)} - 1\right) 
 + H\left(3\Omega_\mathrm{hol}^{(B)} - 3 - \Omega_\mathrm{r}\right) = 0~~.
 \label{new6}
\end{eqnarray}
Using $\Omega_\mathrm{hol}^{(B)} = \kappa^2\rho_\mathrm{hol}^{(B)}/\left(3H^2\right)$ into Eq.(\ref{basic}), we can express the holographic cut-off in terms of 
$\Omega_\mathrm{hol}^{(B)}$ as,
\begin{eqnarray}
 L_\mathrm{B} = \frac{c}{H\sqrt{\Omega_\mathrm{hol}^{(B)}}}~~,
 \label{new7}
\end{eqnarray}
where the form of $L_\mathrm{B}$ in the present context is obtained in Eq.(\ref{rhoB2}) (or Eq.(\ref{rhoBfuture horizon})) in terms of $L_\mathrm{p}$ 
and its derivatives (or $L_\mathrm{f}$ and its derivatives) respectively. 
Differentiation both sides of Eq.(\ref{new7}) (with respect to $t$) immediately leads to,
\begin{eqnarray}
 \frac{1}{L_\mathrm{B}}\frac{dL_\mathrm{B}}{dt} = -\frac{\dot{H}}{H} - \frac{1}{{2\Omega_\mathrm{hol}^{(B)}}}\frac{d\Omega_\mathrm{hol}^{(B)}}{dt}~~.
 \label{new8}
\end{eqnarray}
Using the explicit form of $L_\mathrm{B}$ (from Eq.(\ref{rhoB2}) or Eq.(\ref{rhoBfuture horizon})), we get $\dot{L}_\mathrm{B}/L_\mathrm{B}$ in terms 
of $L_\mathrm{p}$ or $L_\mathrm{f}$ as,
\begin{align}
\frac{1}{L_\mathrm{B}}\frac{dL_\mathrm{B}}{dt} = \frac{\left(\ddot{L}_\mathrm{p}/L_\mathrm{p} - \left(\dot{L}_\mathrm{p}/L_\mathrm{p}\right)^2 
+ \dot{L}_\mathrm{p}/L_\mathrm{p}^2\right)}{\left(\frac{\dot{L}_\mathrm{p}}{L_\mathrm{p}} - \frac{1}{L_\mathrm{p}}\right)}
\left\{\frac{\left(1+\Delta\right)\left[H_1^2\left(\frac{\dot{L}_\mathrm{p}}{L_\mathrm{p}} - \frac{1}{L_\mathrm{p}}\right)^{-2}\right]^{\Delta} - 1}
{1 - \frac{1 + \Delta}{1 - \Delta} \left[H_1^2\left(\frac{\dot{L}_\mathrm{p}}{L_\mathrm{p}} - \frac{1}{L_\mathrm{p}}\right)^{-2}\right]^{\Delta}}\right\}\, ,
\label{new9}
\end{align}
or,
\begin{align}
\frac{1}{L_\mathrm{B}}\frac{dL_\mathrm{B}}{dt} = \frac{\left(\ddot{L}_\mathrm{f}/L_\mathrm{f} - \left(\dot{L}_\mathrm{f}/L_\mathrm{f}\right)^2 
 - \dot{L}_\mathrm{f}/L_\mathrm{f}^2\right)}{\left(\frac{\dot{L}_\mathrm{f}}{L_\mathrm{f}} + \frac{1}{L_\mathrm{f}}\right)}
\left\{\frac{\left(1+\Delta\right)\left[H_1^2\left(\frac{\dot{L}_\mathrm{f}}{L_\mathrm{f}} + \frac{1}{L_\mathrm{f}}\right)^{-2}\right]^{\Delta} - 1}
{1 - \frac{1 + \Delta}{1 - \Delta} \left[H_1^2\left(\frac{\dot{L}_\mathrm{f}}{L_\mathrm{f}}+- \frac{1}{L_\mathrm{f}}\right)^{-2}\right]^{\Delta}}\right\}\, ,
\label{new10}
\end{align}
respectively. Owing to Eq.(\ref{HLL}), we may equivalently express both the Eqs.(\ref{new9}) and (\ref{new10}) as,
\begin{align}
\frac{1}{L_\mathrm{B}}\frac{dL_\mathrm{B}}{dt} = 
\left(\frac{\dot{H}}{H}\right)\left\{\frac{\left(1+\Delta\right)\left(\frac{H_1^2}{H^2}\right)^{\Delta} - 1}
{1 - \frac{1 + \Delta}{1 - \Delta} \left( \frac{H_1^2}{H^2} \right)^{\Delta}}\right\}\, .
\label{new11}
\end{align}
Due to the above expression of $\dot{L}_\mathrm{B}/L_\mathrm{B}$, Eq.(\ref{new8}) turns out to be,
\begin{eqnarray}
\frac{1}{{2\Omega_\mathrm{hol}^{(B)}}}\frac{d\Omega_\mathrm{hol}^{(B)}}{dt} 
+ \frac{\dot{H}}{H}\left\{\frac{\left(1+\Delta\right)\left(\frac{H_1^2}{H^2}\right)^{\Delta} - 1}
{\frac{1 + \Delta}{1 - \Delta} \left( \frac{H_1^2}{H^2} \right)^{\Delta} - 1}\right\} = 0~~.
\label{new12}
\end{eqnarray}
Combining Eqs.(\ref{new6}) and (\ref{new12}) along with a little bit of simplification yields the following equation depicting the evolution of 
Hubble parameter with respect to the redshift factor,
\begin{eqnarray}
 \frac{1}{H(z)}\frac{dH}{dz} = -\frac{\Omega_\mathrm{hol}^{(B)}}{(1+z)}\left(\frac{3}{2} - \frac{3+\Omega_\mathrm{r}}{2\Omega_\mathrm{hol}^{(B)}}\right)
 \left[\frac{1}
 {1 - \Omega_\mathrm{hol}^{(B)}\left\{\frac{\left(1+\Delta\right)\left(\frac{H_1^2}{H^2}\right)^{\Delta} - 1}
{\frac{1 + \Delta}{1 - \Delta} \left( \frac{H_1^2}{H^2} \right)^{\Delta} - 1}\right\}}\right]~~,
\label{new13}
\end{eqnarray}
where $z = -1+\frac{1}{a}$ is the redshift factor and the present scale factor is taken to be unity. Eq.(\ref{new13}) clearly indicates that 
$\frac{dH}{dz}$ becomes zero at $\Omega_\mathrm{hol}^{(B)} = 1 + \Omega_\mathrm{r}/3$. Therefore the present generalized HDE model does not admit 
any turning point in Hubble parameter due to $\Omega_\mathrm{hol}^{(B)} < 1$. This is unlike to the HDE model proposed in \cite{Colgain:2021beg} where the authors 
considered the future horizon to be the holographic cut-off and consequently the respective HDE model allows a turning point in the Hubble parameter evolution. 
However in the present context, the holographic cut-off (corresponds to the Barrow entropic energy density) 
is not same as the future horizon, in particular the form of $L_\mathrm{IR}$ is shown in 
Eq.(\ref{rhoB2}) (or in Eq.(\ref{rhoBfuture horizon})). As a result, the Hubble evolution in the present scenario gets different compared to that of 
in \cite{Colgain:2021beg} and thus the current HDE model does not exhibit any turning point in the Hubble parameter. 

Here we would like to mention that, as demonstrated in \cite{Colgain:2021beg}, the presence of turning point in $H(z)$ may alleviate the Hubble tension. In particular, 
a HDE model can resolve the $H_0$ tension and achieves higher values of $H_0$ (compared to $\Lambda$CDM) 
by having a turning point close enough to $z = 0$ regime. However for the current HDE model (having the cut-off given in Eq.(\ref{rhoB2})), we have found 
that the $H(z)$ does not experience any turning point in its evolution and thus we may argue that this model is unable to resolve the Hubble tension, unless 
some generalizations are incorporated so that it gives an early dark energy (EDE) component. Therefore in order to resolve the $H_0$ tension in the present context, 
the holographic cut-off needs to be modified in such a way that it leads to an EDE and consequently helps to mitigate the Hubble tension issue. 
A phenomenological model of EDE was proposed in \cite{Karwal:2016vyq} where the energy density and pressure of EDE are considered to be,
\begin{eqnarray}
 \rho_\mathrm{ede}&=&\rho_\mathrm{c}\times\frac{\Omega_\mathrm{ede}\left(1 + a_c^6\right)}{a^6 + a_c^6}~~,\nonumber\\
 p_\mathrm{ede}&=&\rho_\mathrm{ede}\left(\frac{a^6 - a_c^6}{a^6 + a_c^6}\right)
 \label{ede1}
\end{eqnarray}
respectively (for other form of EDE, see \cite{Poulin:2018cxd}). In the above expression, 
$\rho_\mathrm{c}$ is the critical density today, $\Omega_\mathrm{ede}$ is the fractional 
energy density of the EDE today and $a_c = 1/(1 + z_c)$ is the critical value of the scale factor at which the EDE shifts 
from early-time behavior to late-time behavior. Here it may be mentioned that the above forms of $\rho_\mathrm{ede}$ and 
$p_\mathrm{ede}$ satisfy the respective conservation relation of EDE. Clearly the equation of state corresponding to the EDE is given by 
$\omega_\mathrm{ede} = \frac{a^6 - a_c^6}{a^6 + a_c^6}$. Therefore at redshifts $z \gg z_c$ 
or equivalently at $a \ll a_c$, $\omega_\mathrm{ede}$ tends to $-1$ and thus the EDE behaves 
like a cosmological constant at early times, similar to a 
slowly rolling scalar field. On other hand, at redshifts $z \ll z_c$ or $a \gg a_c$, we get $\omega_\mathrm{ede}\approx 1$ leading to the late time 
evolution of EDE density as $\rho_\mathrm{ede} \propto \frac{1}{a^6}$ i.e the EDE decays faster than the radiation with the cosmological expansion of the 
universe. The presence of such EDE is able to resolve the Hubble tension as demonstrated in \cite{Karwal:2016vyq} where 
a wide range of the critical redshift parameter has been considered from $z_c = 10$ to $z_c = 10^6$. In particular, 
the authors of \cite{Karwal:2016vyq} found that $\tau$ (reionization optical depth) 
plays an important role in alleviating the Hubble tension. A larger value of $\tau$ leads, with EDE, to a larger best-fit 
value of $H_0$.

At this stage, we determine the modified form of holographic cut-off so that it leads to the aforementioned EDE of Eq.(\ref{ede1}). 
In presence of the EDE, the total holographic energy density in the present context can be written as,
\begin{eqnarray}
 \rho_\mathrm{hol}^{(T)} = \rho_\mathrm{hol}^{(B)} + \rho_\mathrm{c}\times\frac{\Omega_\mathrm{ede}\left(1 + a_c^6\right)}{a^6 + a_c^6}~~,
 \label{ede2}
\end{eqnarray}
where recall, $\rho_\mathrm{hol}^{(B)}$ corresponds to the Barrow entropic energy density shown in Eq.(\ref{rhoB2}) (or in Eq.(\ref{rhoBfuture horizon})). 
By comparing Eq.(\ref{ede2}) and Eq.(\ref{basic}), we obtain the modified holographic cut-off (symbolized by $L_\mathrm{mod}$) that corresponds 
to the dark energy density $\rho_\mathrm{hol}^{(T)}$ as,
\begin{align}
\label{ede3}
\frac{3c^2}{\kappa^2 L^2_\mathrm{mod}} 
= \frac{3}{8\pi G} 
\left( \left( \frac{\dot{L}_\mathrm{p}}{L_\mathrm{p}} - \frac{1}{L_\mathrm{p}} \right)^2 
- \frac{1 + \Delta}{1 - \Delta} H_1^2\left( \frac{\left( \frac{\dot{L}_\mathrm{p}}{L_\mathrm{p}} - \frac{1}{L_\mathrm{p}} \right)^2}
{H_1^2} \right)^{1 - \Delta} + \rho_\mathrm{c}\frac{\Omega_\mathrm{ede}\left(1 + a_c^6\right)}{a^6 + a_c^6} \right) \, , 
\end{align}
in terms of $L_\mathrm{p}$ and its derivatives. Similarly $L_\mathrm{mod}$, in terms of the future horizon and its 
derivatives, is given by,
\begin{align}
\label{ede4}
\frac{3c^2}{\kappa^2 L^2_\mathrm{mod}} = \frac{3}{8\pi G} 
\left( \left( \frac{\dot{L}_\mathrm{f}}{L_\mathrm{f}} + \frac{1}{L_\mathrm{f}} \right)^2 
- \frac{1 + \Delta}{1 - \Delta} H_1^2\left( \frac{\left( \frac{\dot{L}_\mathrm{f}}{L_\mathrm{f}} + \frac{1}{L_\mathrm{f}} \right)^2}
{H_1^2} \right)^{1 - \Delta} + \rho_\mathrm{c}\frac{\Omega_\mathrm{ede}\left(1 + a_c^6\right)}{a^6 + a_c^6} \right) \, .
\end{align}
Thereby the dark energy model consists of Barrow entropic energy density and early dark energy density can be regarded as a candidate of generalized 
holographic dark energy family, with the corresponding holographic cut-off is given by Eq.(\ref{ede3}) (or by Eq.(\ref{ede4})). Clearly the above 
form(s) of $L_\mathrm{mod}$ leads to an early dark energy component from holographic point of view, by which, the Hubble tension can be resolved.

\section{Extended case of Barrow dark energy and its equivalence with generalized holographic dark energy} \label{Sec4}

In this section, we consider the models where the entropy exponent has a running behaviour, in particular varies with 
the cosmological expansion of the universe. In the case 
of entropic DE models other than the Barrow entropic one, the scenario with variable exponent has been discussed in \cite{Nojiri:2019itp,Nojiri:2019skr} 
where the authors claimed that such behaviors may appear because the entropy corresponds to physical degrees of freedom and 
the degrees of freedom depend on the scale as implied by 
the renormalization of a quantum theory. In cosmology, as the energy scale is controlled by the Hubble parameter, we define a dimensionless 
variable as $x = H_1^2/H^2$, where recall, $H_1^2 = 4\pi/A_0$. If we use this extended formalism (where the exponent of the respective 
entropy function varies) to the Barrow entropy, then the modified Barrow entropy function can be written as,
\begin{align}
S_B = \frac{A_0}{4G}\left(\frac{A}{A_0}\right)^{1+\Delta(x)}\, .
\label{extended-entropy}
\end{align}
Note that this is qualitatively similar to generalized Tsallis entropic DE introduced in \cite{Nojiri:2019skr} 
where also time-dependence of entropy functional is explicit. Using $A = 4\pi r_h^2$, we determine from Eq.(\ref{extended-entropy}),
\begin{align}
\frac{dS_B}{dt}=&\frac{\partial S}{\partial A}\frac{dA}{dt} + \frac{\partial S}{\partial x}\frac{dx}{dt}\nonumber\\
=&-\frac{1}{4G}\left(\frac{8\pi}{H^3}\right)\left(\frac{H_1^2}{H^2}\right)^{\Delta(x)}
\left\{\left(1 + \Delta(x)\right) + \frac{H_1^2}{H^2}\ln{\left(\frac{H_1^2}{H^2}\right)}\Delta'(x)\right\}\dot{H}\, .\nonumber\\
\label{ext-time derivative entropy}
\end{align}
By applying the first law of thermodynamics, we obtain the second FLRW equation for this extended scenario of the Barrow entropy as,
\begin{align}
\left\{\left(1+\Delta(x)\right) + \frac{H_1^2}{H^2}\ln{\left(\frac{H_1^2}{H^2}\right)}\Delta'(x)\right\}
\left(\frac{H_1^2}{H^2}\right)^{\Delta(x)}\dot{H} 
= -4\pi G\left(\rho + p\right)\, ,
\label{ext-FLRW1}
\end{align}
where $\rho$ and $p$ are the energy density and pressure of the ordinary matter contents. It may be observed that the running 
behaviour of $\Delta(x)$ affects the FLRW equation compared to the case where the exponent remains constant (see Eq.~(\ref{Barrow-FLRW1})). 
Using the conservation relation of $\rho$, one may integrate the above equation to get the first FLRW equation as,
\begin{equation}
\label{ext-FLRW2}
\left. - H_1^2 \left\{ x^{\Delta(x) - 1} + 2 \int^x dx~x^{\Delta(x) -2} 
\right\}
\right|_{x=\frac{H_1^2}{H^2}}= \frac{8\pi G}{3} \rho + \frac{\Lambda}{3} \, .
\end{equation}
To get the above expression, we use $H\frac{dH}{dx} = -\frac{H_1^2}{2x^2}$. Eqs.~(\ref{ext-FLRW1}) and (\ref{ext-FLRW2}) are the modified FLRW equations 
for the Barrow entropic energy density with variable exponent. At this stage, if we introduce $\rho_\mathrm{B}$ and $p_\mathrm{B}$ as,
\begin{equation}
\label{ext-Barrow energy density}
\rho_\mathrm{B} = \frac{3}{8\pi G} 
\left( H^2 + \left. H_1^2 \left\{ x^{\Delta(x) - 1} + 2 \int^x dx~x^{\Delta(x) -2} 
\right\}\right|_{x=\frac{H_1^2}{H^2}} \right)\, ,
\end{equation}
and
\begin{align}
p_\mathrm{B} = \frac{\dot{H}}{4\pi G}\left\{x^{\Delta(x)}\left(1+\Delta(x)\right) + x^{1+\Delta(x)}\ln{(x)}\Delta'(x) - 1\right\} - \rho_\mathrm{B}
\label{ext-Barrow pressure}
\end{align}
respectively, then the FLRW equations (i.e. Eqs.~(\ref{ext-FLRW1}) and (\ref{ext-FLRW2})) can be equivalently written as,
\begin{align}
\dot{H}=&-4\pi G\left[\left(\rho + p\right) + \left(\rho_\mathrm{B} + p_\mathrm{B}\right)\right]\, ,\nonumber\\
H^2=&\frac{8\pi G}{3}\left(\rho + \rho_\mathrm{B}\right) + \frac{\Lambda}{3}\, .
\label{ext-modified FLRW}
\end{align}
The above equations clearly indicate that $\rho_\mathrm{B}$ and $p_\mathrm{B}$ represent the energy density and pressure corresponding to the 
Barrow entropy with varying exponent. In order to get an explicit form of $\rho_\mathrm{B}$ (and $p_\mathrm{B}$), one needs a functional form of 
$\Delta(x)$ to perform the integration of Eq.~(\ref{ext-Barrow energy density}). As mentioned earlier, the presence of the varying exponent 
significantly modifies the FLRW equations compared to that of the original one. Clearly if $\Delta(x)$ becomes constant at late time, then 
the outcome from this extended scenario will match with the previously discussed the Barrow dark energy model where the entropy exponent is considered to be 
a constant. However with the varying exponent, the Barrow entropic scenario may lead to more interesting and richer phenomenology compared 
to the previous case where the exponent is considered to be constant. 
In particular, it can lead to an unified scenario of early inflation and late dark energy epoch with an intermediate deceleration era, 
if $\Delta(x)$ is chosen in such a way that at high and low energy scales it gets the values away from the standard value $= 1$, 
while at intermediate scales it remains close to unity. However the cosmology for the Barrow entropy with varying exponent has not been discussed 
in earlier literatures, and thus it will be important to discuss the possible effects of $\rho_\mathrm{B}$ (of Eq.~(\ref{ext-Barrow energy density})) 
in driving the inflation, the late time acceleration and even a non-singular bounce. 
This is out of the scope of the present work and is expected to study elsewhere.

Coming back to Eqs.~(\ref{ext-Barrow energy density}) and (\ref{ext-Barrow pressure}), we can define an equation of state parameter as,
\begin{align}
\omega_\mathrm{B} = \frac{p_\mathrm{B}}{\rho_\mathrm{B}} 
= -1 + \frac{2\dot{H}}{3H^2}\left\{\frac{x^{\Delta(x)}\left(1 + \Delta(x)\right) + x^{1+\Delta(x)}\ln{(x)}\Delta'(x) - 1}
{1 + x^{\Delta(x)} + 2x\int^{x}dx~x^{\Delta(x)-2}}\right\}\, .
\label{ext-eosB}
\end{align}
This is the effective EoS parameter for the extended Barrow entropy scenario where the exponent varies with the cosmological evolution of the universe. 

Having set the stage, we now establish the equivalence between the extended version of the Barrow energy density and the generalized holographic 
energy density. For this purpose, we use Eq.~(\ref{ext-Barrow energy density}) which immediately leads to the 
equivalent holographic cut-off corresponding to the extended Barrow entropic scenario as,
\begin{align}
\label{ext-equiv-cut-off1}
\frac{3c^2}{\kappa^2 L^2_\mathrm{B}} 
= \frac{3}{8\pi G} 
\left( \left( \frac{\dot{L}_\mathrm{p}}{L_\mathrm{p}} - \frac{1}{L_\mathrm{p}} \right)^{2} + \left. H_1^2 
\left\{ x^{\Delta(x) - 1} + 2 \int^x dx x^{\Delta(x) -2} 
\right\}\right|_{x=H_1^2\left( \frac{\dot{L}_\mathrm{p}}{L_\mathrm{p}} - \frac{1}{L_\mathrm{p}} \right)^{-2}} \right)\, ,
\end{align}
in terms of $L_\mathrm{p}$ and its derivative, or,
\begin{align}
\label{ext-equiv-cut-off2}
\frac{3c^2}{\kappa^2 L^2_\mathrm{B}} = \frac{3}{8\pi G} 
\left( \left( \frac{\dot{L}_\mathrm{f}}{L_\mathrm{f}} + \frac{1}{L_\mathrm{f}} \right)^{2} + \left. H_1^2 
\left\{ x^{\Delta(x) - 1} + 2 \int^x dx x^{\Delta(x) -2} 
\right\}\right|_{x=H_1^2\left( \frac{\dot{L}_\mathrm{f}}{L_\mathrm{f}} + \frac{1}{L_\mathrm{f}} \right)^{-2}} \right)\, ,
\end{align}
in terms of $L_\mathrm{f}$ and its derivative, respectively. Consequently, the holographic EoS parameter, in particular for 
$\rho_\mathrm{hol} \propto 1/L_\mathrm{B}^2$, is given by,
\begin{align}
W_\mathrm{hol}^\mathrm{(B)} = -1 + \left(\frac{2}{3HL_\mathrm{B}}\right)\frac{dL_\mathrm{B}}{dt}\, ,
\label{ext-eos-holographic}
\end{align}
with $L_\mathrm{B}$ is shown in Eq.~(\ref{ext-equiv-cut-off1}) (or Eq.~(\ref{ext-equiv-cut-off2})). Using such expression of $L_\mathrm{B}$ along with 
Eq.~(\ref{HLL}), we determine,
\begin{align}
\frac{3c^2}{\kappa^2L_\mathrm{B}^2}\left(\frac{2\dot{L}_\mathrm{B}}{3L_\mathrm{B}}\right) 
= \frac{1}{4\pi G}\left\{\left(1+\Delta(x)\right)x^{\Delta(x)} + x^{1+\Delta(x)}\ln{(x)}\Delta'(x) - 1\right\}H\dot{H}\, .
\label{ext-eos-equiv}
\end{align}
By plugging the above expression into Eq.~(\ref{ext-eos-holographic}) and comparing with Eq.~(\ref{ext-eosB}) immediately lead to, 
\begin{align}
W_\mathrm{hol}^\mathrm{(B)} = \omega_\mathrm{B}\, .
\end{align}
Therefore even for the case where the exponent varies with the cosmological evolution, the Barrow entropic energy density as well as 
the Barrow EoS parameter seem to be equivalent with that of in the generalized holographic scenario, with the equivalent NO cut-off 
is given in Eq.~(\ref{ext-equiv-cut-off1}) or Eq.~(\ref{ext-equiv-cut-off2}). Interestingly, such cut-off is found to depend up-to first 
derivative of $L_\mathrm{p}$ and $L_\mathrm{f}$ respectively.\\

Before concluding, here it may be mentioned that it will be an interesting avenue to apply the present method ( used to establish 
the equivalence between the Barrow entropic dark energy and the generalized HDE ) for various early universe scenario. 
In particular, recently an unified cosmological model has been proposed from a non-singular bounce to the dark energy era 
with an intermediate matter dominated epoch, and moreover the primordial 
as well as the dark energy observable quantities are found to be compatible with the latest Planck constraints \cite{Odintsov:2020zct,Odintsov:2021yva}. 
The equivalence of such unified scenario with the generalized HDE will be examined elsewhere. 

\section{Conclusion}

The holographic principle along with dimensional analysis leads to the formulation of holographic dark energy (HDE) 
where the dark energy density gets proportional to the inverse squared of the infrared cut-off. 
It may be mentioned that instead of adding a term in the Lagrangian, the HDE 
is based on the holographic principle, due to which, the HDE becomes significantly different than the usual dark energy (DE) models 
where the dark energy density comes from some suitable scalar field or by higher curvature term(s) present in the Lagrangian. Usually, 
the infrared cut-off is taken as the particle horizon ($L_\mathrm{p}$) or the future horizon ($L_\mathrm{f}$). However, 
in \cite{Nojiri:2005pu}, a generalized HDE has been proposed where the cut-off ($L_\mathrm{IR}$) is generalized to depend on 
$L_\mathrm{IR} = L_\mathrm{IR} \left(L_\mathrm{p}, \dot L_\mathrm{p}, 
\ddot L_\mathrm{p}, \cdots, L_\mathrm{f}, \dot L_\mathrm{f}, \cdots, a\right)$. Such 
form of $L_\mathrm{IR}$ evidently allows a more rich phenomenology to the generalized HDE.

Based on the generalized holographic formalism, we have showed that the Barrow entropic dark energy is equivalent with the generalized HDE with a suitable 
form of the holographic cut-off. In particular, the holographic cut-off is determined either in terms of particle horizon and its derivative 
(with respect to cosmic time) or in terms of future horizon and its derivative. Such equivalence between the Barrow entropic dark energy and the generalized HDE 
has been established for two cases: (1) when the Barrow exponent is a constant and (2) when the exponent is allowed to vary with the cosmological expansion 
of the universe. In the former case, the equivalent holographic cut-off is found to depend up-to first order derivative of the particle horizon 
(or the future horizon). The second case. i.e. when the exponent varies with the cosmological evolution, is important to study 
because the entropy corresponds to physical degrees of freedom and the degrees of freedom depend on the scale as implied by the renormalization of a 
quantum theory. Such extended formalism of entropic dark energy has been applied earlier, however for the Tsallis and the Renyi DE models. 
Unlike to earlier literatures, in the present paper, we applied the extended formalism to the Barrow entropic dark energy model 
and established its equivalence with the generalized HDE. Furthermore, it deserves mentioning that for both the cases (whether the exponent is constant 
or varies), we determined the equation of state (EoS) parameter both for the Barrow entropic dark energy and the corresponding generalized HDE models, 
with the EoS parameter are symbolized by $\omega_\mathrm{B}$ and $W_\mathrm{hol}^\mathrm{(B)}$ respectively. As a result, it turns out that 
$\omega_\mathrm{B} = W_\mathrm{hol}^\mathrm{(B)}$, which further ensures the equivalence between the Barrow entropic dark energy and the generalized HDE.

In summary, the Barrow entropic dark energy model is found to be equivalent with the generalized HDE, with the corresponding cut-off 
is determined in terms of particle horizon and its derivative or in terms of future horizon and its derivative. 
However, the choice of fundamental viable cut-off is still a debatable topic and an active area of research. 
The comparison of such cut-offs for realistic description of the universe evolution may help in 
better understanding of holographic principle.

\section*{Acknowledgments}
This work is supported by the JSPS Grant-in-Aid for Scientific Research (C)
No. 18K03615 (S.N.) and by MINECO (Spain), project PID2019-104397GB-I00 (SDO).

\end{document}